# Rebuttal to Berger et al., TOPLAS 2019

Baishakhi Ray, Prem Devanbu, Vladimir Filkov
rayb@cs.columbia.edu, devanbu@cs.ucdavis.edu, filkov@cs.ucdavis.edu

"The first principle is that you must not fool yourself — and you are the easiest person to fool."
— Richard Feynman, 1974

**Abstract**

Berger et al., published in TOPLAS 2019[1], is a critique of our 2014 FSE conference abstract and its archival version, the 2017 CACM paper: *A Large-Scale Study of Programming Languages and Code Quality in Github*[2]. In their paper Berger et al. make academic claims about the veracity of our work. Here, we respond to their technical and scientific critiques aimed at our work, attempting to stick with scientific discourse. We find that Berger et al. largely replicated our results, and agree with us in their conclusion: that the effects (in a statistical sense) found in the data are small, and should be taken with caution, and that it is possible that an absence of effect is the correct interpretation. Thus, our CACM paper's conclusions still hold, even more so now that they've been reproduced, and our paper is eminently citable.

**Introduction**

Replication studies are relatively rare. Scientists wish that weren't so, as independent replication is at the core of what good science should be. Unfortunately, it seems funding for replications is less available than funding for doing new things. So, we were delighted to hear that Dr. Jan Vitek and his student were interested in doing a replication of our paper *A Large-Scale Study of Programming Languages and Code Quality in Github*. We happily provided our data from an earlier version of our paper, presented at the FSE 2014 conference, so they could do so. The paper by Berger, Hollenbeck, Maj, Vitek, and Vitek, published in TOPLAS 2019, *On the Impact of Programming Languages on Code Quality: A Reproduction Study*, is their final product: a reproduction and critique of our CACM 2017 paper, in which they also refer to our paper's earlier, conference version from FSE 2014.

To avoid confusion, we note that our CACM 2017 paper is an improved, archival/journal version of our FSE 2014 conference paper of the same title. That conference paper reported on preliminary results. Upon it being invited as a research highlight, after being recommended by the FSE 2014 conference program committee chairs, it underwent multiple review rounds at CACM. The CACM 2017 version is much improved over the FSE 2014 version and completely supersedes it. It is standard practice in Computer Science to have conference paper abstracts extended/improved and published in an archival form in a journal. Once the journal version is published, other papers begin to cite it and stop citing the conference version. Berger et al. refer to results from both our preliminary, FSE paper and our final CACM paper in their TOPLAS paper comparisons, which creates confusion.

---

[1] Berger et al., On the Impact of Programming Languages on Code Quality: A Reproduction Study, TOPLAS 2019, https://dl.acm.org/citation.cfm?id=3340571
[2] https://cacm.acm.org/magazines/2017/10/221326-a-large-scale-study-of-programming-languages-and-code-quality-in-github/fulltext#comments

In the next sections, we give a point-by-point rebuttal to all the critical points Berger et al. raise in their TOPLAS paper and multiple talks arising from it. In summary:

- In CACM 2017 we reported a *weak* association between programming languages and bug proneness, and warned readers (in the abstract!) that *"However, we caution the reader that even these modest effects might quite possibly be due to other, intangible process factors, for example, the preference of certain personality types for functional, static languages that disallow type confusion"*. Yet, Burger et al. make claims to the contrary.
- In their paper, Berger et al. essentially repeated parts of our work and obtain [mostly confirmatory results](): both they and we conclude on a negative note, that the small effects should be taken with caution, and that it is possible that an absence of effect is the correct interpretation. In spite of this concurrence in the results of both studies, in a surprising contrast with their results, Berger et al. claim that our study is beyond repair, and that theirs proves it wrong.
- After careful reading of their paper, we find that the claimed differences between their central results, given in their Table 6, and ours in CACM, mostly arise due to Berger et al.'s decision to derive their conclusions from the results of an overly conservative statistical approach, the Bonferroni correction, in spite of (1) being aware[3] that it is not appropriate, and that using it would result in fewer, if any, statistically significant results, and in spite of (2) them having performed a more appropriate correction, the FDR, which they decided not to use in the interpretation of their final results).
- They have a lot of criticisms of our approach; as we argue below a great many of these criticisms are either oversights on their part, misinterpretations of our and their results, and/or are matters of opinion inappropriately presented as fact.
- We also found a few interesting methodological turns in their paper, a smaller subset of which we agree are worth exploring further.
- Both they and we conclude on the same note: that the small effects exist, but should be taken with caution, and that it is possible that an absence of effect is the correct interpretation. <u>In conclusion, it's clear to us that the findings of our CACM 2017 paper still stand.</u>

---

[3] Indeed one of their co-authors, Dr. O. Vitek, a statistician, has repeatedly used the FDR over Bonferroni in her own writings

**Point-by-Point Rebuttal**

We will use **FSE14** for our FSE 2014 paper (preliminary results), **CACM17** for our CACM 2017 improved journal version, and **TOPLAS19** for their TOPLAS 2019 paper. We will also use their TOPLAS terms: **repetition** for their rerunning of our scripts on our data[4], and **reanalysis** for their regathering data and cleaning up of their and our data and using their methods to analyze the newly cleaned data.

We also annotate our points with:

- **#miss** (for misinterpretations of ours or their findings),
- **#opinion** (for debatable opinions that are nevertheless stated as facts, in their claims),
- **#future** (for valid points for future work).

Below, we follow the enumeration of the claims from the Curry On! Talk by Dr. J. Vitek (the "7 Sins").

**Claim 1:** The size of projects are too different (some have millions of lines of code, others just tens of lines), yet in **CACM17** the authors don't normalize for the number of commits. Hence the results may be wrong!

Answer: **#miss** Demonstrably false. We do take into account the size of the projects/number of commits, using the variables Log(size) and Log(commits), see **CACM17**, Table 6, and the description in the text.

**Claim 2:** There seem to be uncontrolled effects, hence the results may be wrong!

Answer: **#miss** This is a rather extreme position, which prima facie appears to reject large segments of empirical science writ large. We controlled for a number of project-specific effects, including age, size, number of developers, and number of commits; all this is based on existing findings in Software Engineering research. There will always be uncontrolled-for effects in any empirical study, especially when studying such a complex phenomenon as commit bugginess. Not doing empirical work "because we can't know everything" is a self-reinforcing argument for preserving ignorance. Experimental methods aim to chip away at the boundaries of knowledge. Our specific goal was to identify effects (in a statistical sense) of individual languages on the outcome, while controlling for factors accepted as relevant in the field, at the time the experiment was done. As in every scientific endeavour, knowledge and methods improve with time. Such methods are routinely and effectively used, *e.g.,* in public health policy, economics, climate science, management science, and in the software engineering industry.

**Claim 3:** Duplicated commits exist in the data, to the tune of < 2%. Hence the results may be wrong!

Answer: **#opinion** We did not look for duplicated commits but are not surprised that such commits may exist: notably they amount to less than 2% overall! **TOPLAS19** doesn't provide any evidence that the duplicated commits are anyhow biased, i.e., either more or less buggy than the 98+% non-duplicated ones. Thus, it is very unlikely that this affects our results. In fact, the results in **TOPLAS19** mostly confirm that it doesn't, see claim 6 below.

---

[4] The repetition was done on some of our data and with some of our scripts, which we were happy to share with one of the authors, Dr. Jan Vitek, and a student, upon their request in 2017.

**Claim 4:** There is missing data, about 20% over all projects, and up to 80% for some projects. Hence, the results may be wrong!

Answer: **#miss** We assert that this statement, and it's, Figure 3 in **TOPLAS19,** are highly disingenuous. Fig 3 shows all the bar plots on the same chart, and truncates (shortens) the 80% bar (for Perl), thus most bars look comparable to it, appearing as if for most languages we have lost most commits. In fact, looking at the figure's y-axis units, we see that there is only one language for which the discrepancy is at 80%, Perl. For all others the discrepancy is below 15%, and for most way below. This is classical data misrepresentation, usually only seen in tendentious political debates.

We can offer some possible reasons for the missing data (we assume that Berger et al. were careful to account for GitHub changes over time between their and our pubs; and we assume Berger et al. were careful to only look for commits with dates before our data gathering date, in 2013). We associated a language to a committed file based on the "primary_extension" of the studied languages, as per GitHub Linguist[5]. Thus, we disregarded file-commits that do not have the primary extensions of the studied languages. This resulted in a conservative collection of commits, but one aimed at lowering mismatching of languages, i.e., Type I errors.

For Perl, specifically, we found that a few large projects used a .pm extension (a secondary extension) for files with most commits, which we disregarded as we only looked at .pl files, the primary Perl extension. As there is no reason to suspect any bias in the data, a 20% sampling of the data should be representative of the whole set.

In any case, the relevance of the missing data is somewhat moot since **TOPLAS19** confirmed our results for most languages (7 out of 9), see point 6. below.

**Claim 5:** There are Typescript misidentification issues in **FSE14.** There are bug-fix misidentification issues in the project V8. Thus, other problems may exist!

Answer: **#miss** On TypeScript: This is an unnecessary misrepresentation of our work. We were the first to correct our work immediately after noticing this very issue. In **FSE14** we relied on GitHub's language classification, specifically on Linguist. We recognized issues with Linguist's identification of TypeScript files very soon after the publication of **FSE14**. We reran the analyses, and corrected the results in our paper pdf's on our sites online[6]. We also wrote to the FSE 2014 PC chairs with a corrected version. The correction resulted in many fewer projects being identified as TypeScript projects, and consequently in non-significant findings for the association of TypeScript with defects. The **CACM17** paper uses the corrected TypeScript data, and Berger et al. were aware of that, yet they chose to compare to the version of our **FSE14** paper that has a mistake in it. This seems unnecessarily tendentious.

---

[5] https://github.com/github/linguist/blob/17d54d61b423c009e96daca4c0e7f3e9d9df388f/lib/linguist/languages.yml

**#opinion #future** On V8: The V8 example is interesting from a software engineering perspective. Yes, a significant amount of the JavaScript code in V8 is test code. However, coding (and fixing) test files is a big part of what developers do, and thus, there are good experimental reasons to include them. Moreover, of the .js files modified in commits, only 22% are test files that have line additions without deletions. One view might be that this number is an upper bound to how much of our V8 Javascript data can be seen as tainted by mere addition (rather than maintenance) of tests. But even that is an overestimate as many bugs in the wild are fixed by adding lines only, also known as *bugs of omission*[7]. Without manual inspection one cannot make a definitive determination either way.

**Claim 6:** Critical statistical issues found during repetition of study. Hence, **FSE14** and **CACM17** can't be right.

Answer: Unfortunately, **TOPLAS19** is misleading when it comes to what the actual results are, both theirs and ours. The **FSE14** and **CACM17** abstracts and conclusions clearly state that though we found statistically significant results in some of our RQs (not all), we very carefully discussed that those are very small and are to be taken with caution. In both **FSE14** and **CACM17** we were clear that we are talking of associations and correlations, not causality[8]. The word *effect* is often used in statistics to indicate the statistical effect of a predictor variable on the dependent variable in regression modeling.

**#miss** Although they claim otherwise, they essentially reproduced our RQ1 results from **FSE14** (compare Table 2a and 2c, **TOPLAS19,** reproduced below).

---

[6] See dated, archived version from 2015, https://web.archive.org/web/20150219105607/http://web.cs.ucdavis.edu/~filkov/papers/
[7] https://www.johndcook.com/blog/2010/01/12/software-sins-of-omission/ , also see this patch from Defects4J http://program-repair.org/defects4j-dissection/#!/bug/Closure/12
[8] Berger et al. object to the world's interpretation of our work, which is not something that should be addressed to us. They disingenuously attempted to pin this on us by erroneously quote mining and/or misquoting us in the **TOPLAS19** intro, which a quick perusal of the original texts can quickly confirm. In spite of us communicating to Berger et al. that they have misquoted us in an earlier, arxiv version of their paper, they apparently did not correct those misquotes in their final **TOPLAS19**.

Table 2. Negative Binomial Regression for Languages (Gray Indicates Disagreement with the Conclusion of the Original Work)

|  | Original Authors | | | | Repetition | |
|---|---|---|---|---|---|---|
|  | (a) FSE [26] | | (b) CACM [25] | | (c) | |
|  | Coef | P-val | Coef | P-val | Coef | P-val |
| Intercept | −1.93 | <0.001 | −2.04 | <0.001 | −1.8 | <0.001 |
| log commits | 2.26 | <0.001 | 0.96 | <0.001 | 0.97 | <0.001 |
| log age | 0.11 | <0.01 | 0.06 | <0.001 | 0.03 | 0.03 |
| log size | 0.05 | <0.05 | 0.04 | <0.001 | 0.02 | <0.05 |
| log devs | 0.16 | <0.001 | 0.06 | <0.001 | 0.07 | <0.001 |
| C | 0.15 | <0.001 | 0.11 | <0.01 | 0.16 | <0.001 |
| C++ | 0.23 | <0.001 | 0.18 | <0.001 | 0.22 | <0.001 |
| C# | 0.03 | – | −0.02 | – | 0.03 | 0.602 |
| Objective-C | 0.18 | <0.001 | 0.15 | <0.01 | 0.17 | 0.001 |
| Go | −0.08 | – | −0.11 | – | −0.11 | 0.086 |
| Java | −0.01 | – | −0.06 | – | −0.02 | 0.61 |
| Coffeescript | −0.07 | – | 0.06 | – | 0.05 | 0.325 |
| Javascript | 0.06 | <0.01 | 0.03 | – | 0.07 | <0.01 |
| Typescript | −0.43 | <0.001 | 0.15 | – | −0.41 | <0.001 |
| Ruby | −0.15 | <0.05 | −0.13 | <0.01 | −0.13 | <0.05 |
| Php | 0.15 | <0.001 | 0.1 | <0.05 | 0.13 | 0.009 |
| Python | 0.1 | <0.01 | 0.08 | <0.05 | 0.1 | <0.01 |
| Perl | −0.15 | – | −0.12 | – | −0.11 | 0.218 |
| Clojure | −0.29 | <0.001 | −0.3 | <0.001 | −0.31 | <0.001 |
| Erlang | 0 | – | −0.03 | – | 0 | 1 |
| Haskell | −0.23 | <0.001 | −0.26 | <0.001 | −0.24 | <0.001 |
| Scala | −0.28 | <0.001 | −0.24 | <0.001 | −0.22 | <0.001 |

We believe the apparent small differences are attributable to changes our artifacts were going through as we were transitioning the **FSE14** artifact to the **CACM17** artifact (close in time to when we shared the **FSE14** artifact with Dr. J. Vitek and his student in 2017).

**#miss** Although they claim otherwise, they reproduced our RQ2 results (compare the bottom rows of Table 4a and 4b, **TOPLAS19**, reproduced below).

Table 4. Negative Binomial Regression for Language Classes

|  | (a) Original | | (b) Repetition | | (c) Reclassification | |
|---|---|---|---|---|---|---|
|  | Coef | P-val | Coef | P-val | Coef | P-val |
| Intercept | −2.13 | <0.001 | −2.14 | <0.001 | −1.85 | <0.001 |
| log age | 0.07 | <0.001 | 0.15 | <0.001 | 0.05 | 0.003 |
| log size | 0.05 | <0.001 | 0.05 | <0.001 | 0.01 | 0.552 |
| log devs | 0.07 | <0.001 | 0.15 | <0.001 | 0.07 | <0.001 |
| log commits | 0.96 | <0.001 | 2.19 | <0.001 | 1 | <0.001 |
| Fun Sta Str Man | −0.25 | <0.001 | −0.25 | <0.001 | −0.27 | <0.001 |
| Pro Sta Str Man | −0.06 | <0.05 | −0.06 | 0.039 | −0.03 | 0.24 |
| Pro Sta Wea Unm | 0.14 | <0.001 | 0.14 | <0.001 | 0.19 | 0 |
| Scr Dyn Wea Man | 0.04 | <0.05 | 0.04 | 0.018 | 0 | 0.86 |
| Fun Dyn Str Man | −0.17 | <0.001 | −0.17 | <0.001 | – | – |
| Scr Dyn Str Man | 0.001 | – | 0 | 0.906 | – | – |
| Fun Dyn Wea Man | – | – | – | – | −0.18 | <0.001 |

Language classes are combined procedural (Pro), functional (Fun), scripting (Scr), dynamic (Dyn), static (Sta), strong (Str), weak (Wea), managed (Man), and unmanaged (Unm). Rows marked – have no observation.

**#miss** RQ2+reclassification: Confusingly, as part of their **repetition** of our RQ2, they changed our classification of programming languages because they disagreed with it, see item below on our take of their classification. After the reclassification they got *qualitatively* the same results as ours: comparing tables 4c in **TOPLAS19** (above) and Table 7 in **CACM17** (below), it is clear they both imply that the functional language categories are associated with (very slightly) fewer bugs, and that those findings are statistically significant.

Table 7: **Functional languages have a smaller relationship to defects than other language classes where as procedural languages are either greater than average or similar to the average. Language classes are coded with weighted effects coding so each language is compared to the grand mean.** *AIC=10419, Deviance=1132, Num. obs.=1067*

| Defective Commits | | |
|---|---|---|
| (Intercept) | $-2.13$ | $(0.10)^{***}$ |
| log commits | $0.96$ | $(0.01)^{***}$ |
| log age | $0.07$ | $(0.01)^{***}$ |
| log size | $0.05$ | $(0.01)^{***}$ |
| log devs | $0.07$ | $(0.01)^{***}$ |
| `Functional-Static-Strong-Managed` | $-0.25$ | $(0.04)^{***}$ |
| `Functional-Dynamic-Strong-Managed` | $-0.17$ | $(0.04)^{***}$ |
| `Proc-Static-Strong-Managed` | $-0.06$ | $(0.03)^{*}$ |
| `Script-Dynamic-Strong-Managed` | $0.001$ | $(0.03)$ |
| `Script-Dynamic-Weak-Managed` | $0.04$ | $(0.02)^{*}$ |
| `Proc-Static-Weak-Unmanaged` | $0.14$ | $(0.02)^{***}$ |

$^{***}p < 0.001, ^{**}p < 0.01, ^{*}p < 0.05$

Alas, for unknown reasons Berger et al. instead of comparing the implications qualitatively, like we did above, they compared their and our regression model coefficients and counted disagreements. That is unsound, and goes against standard statistical practice for regression models with different predictor variables[9]!

We note that this should have been called a *reanalysis*. Berger et al. called it **repetition**, and included it in the **TOPLAS19** repetition section, thereby making it appear as if something was wrong with the scripts/data we shared with them (repetition is a test of the original approach without any modification).

**#opinion** On their reclassification: debate about language classification into strong/weak, dynamic/static, etc., is ongoing in SE/PL, and for some languages it is subjective, as we noted in **CACM17**. Our **CACM17** language classification was a result of many discussions, including a discussion with CACM reviewers and editors. Much of our reasoning is given in **CACM17**. Contrary to what **TOPLAS19** authors state, Scala and Objective-C are not misclassified in **CACM17**, as there are independent resources and opinions classifying them the way we did. E.g., https://en.wikipedia.org/wiki/Objective-C, and https://en.wikipedia.org/wiki/Scala_(programming_language). They would be correct, though, if they had said that our classification is not the only one possible. We note that they compared their results to **FSE14**, the superseded study. In the latter, **CACM17,** we revised our classification of languages, e.g. with TypeScript.

---

[9] Their model has different variables than ours, they derived two categories from one of ours. Thus, the models are not directly comparable, only the implications of those models are comparable. Comparing coefficients across models with different variables is not a standard statistical practice, and can result in unsound conclusions, especially when the new categories result from splitting up old categories, e.g., see Ecological fallacy, https://en.wikipedia.org/wiki/Ecological_fallacy, and Simpson's paradox, https://en.wikipedia.org/wiki/Simpson%27s_paradox.

**#miss** They reproduce our RQ3 results, and they acknowledge this in Sect. 3.2.3. They implemented their own methods for RQ3, different than ours (in the data and scripts package we sent the **TOPLAS19** authors, by mistake we had omitted the scripts to reproduce our RQ3 and RQ4; they did not follow up to ask us for them). Thus, again, this is a reanalysis, so it does not belong in the repetition section. From their results, they conclude the same as we do from ours in **CACM17**: no evidence is found of a correlation between domain and defect proneness. Thus, this is a confirmatory reproduction study of ours.

They did not perform a repetition of our RQ4 as they did not have our scripts, see previous paragraph.

Repetition summary: Berger et al. reproduced our findings on all RQs they attempted: RQ1, RQ2 (using our and their classification), and RQ3 (using their own methods). So, it is incorrect to claim their findings were at odds with ours qualitatively. It is also a misrepresentation to state that reproduction was not possible because a script was missing in the artifact we shared with them. It is also misleading to claim a reproduction was a repetition when it was a reanalysis, especially after spending a significant amount of space in their paper detailing the differences between the two.

**Claim 7:** Critical statistical issues found during reanalysis of study. Hence, **FSE14** and **CACM17** can't be right.

Answer: This is where **TOPLAS19** is most misleading, on multiple accounts. First, their **reanalysis** is only an RQ1 reanalysis. They did not do a reanalysis of our RQ2-RQ4. They gathered their own data, for the same projects we did. For various reasons they couldn't mine all the projects we did. They also could get more data for some projects than we did. This is discussed in Claim 4., above.

**#miss** They compared their results to our preliminary results in **FSE14**, instead of our **CACM17** results, which are slightly different, and the latter supersedes the former. The **TOPLAS19** authors were aware that our **CACM17** was the definitive version (e.g., see the Introduction of **TOPLAS19**), yet chose to compare to **FSE14**.

**#opinion** They correct p-values for multiple hypothesis testing, though whether to correct or not in such a way is a matter of debate[10], especially p-values of coefficients within the same regression model. Still, we recognize that some may argue that a balanced correction like the false discovery rate is appropriate. After performing such a correction in **TOPLAS19,** only 2 of our weakly significant RQ1 findings in **CACM17** become non-significant.

---

[10] A lot has been said about whether a p-value adjustment is, in general, needed or even appropriate, e.g.: http://www.stat.columbia.edu/~gelman/research/published/multiple2f.pdf.

**#miss** Though they claim their results are different than ours, their model coefficients in Table 6b (see above) show very close agreement to ours in Table 6 of **CACM17**. Moreover, their FDR corrected p-values in Table 6c, agree with our Table 6 results in **CACM17** for 7 out of the 9 languages we found significant! Even in the two cases where the two disagree, for PHP and Python, our results are at 0.05 statistical significance, and theirs at 0.075, a small difference. We produce below a table fusing parts of Table 6 from **CACM17** and Table 6c from **TOPLAS19**, since their Table 6 omits our **CACM17** results and only presents the **FSE14** results.

|  | CACM2017 Table 6 | TOPLAS 2019 Table 6c (p-values) | | |
|---|---|---|---|---|
|  | Coef. (Std. Err.) | FDR | Bonf | |
|  | −2.04 (0.11)*** | − | − | |
|  | 0.06 (0.02)*** | − | − | |
|  | 0.04 (0.01)*** | − | − | |
|  | 0.06 (0.01)*** | − | − | |
|  | 0.96 (0.01)*** | − | − | |
| C | 0.11 (0.04)** | 0.017 | 0.118 | Both Significant |
| C++ | 0.18 (0.04)*** | <0.01 | <0.01 | |
| C# | −0.02 (0.05) | 0.85 | 1 | |
| Objective-C | 0.15 (0.05)** | 0.013 | 0.079 | Both Significant |
| Go | −0.11 (0.06) | 0.157 | 1 | |
| Java | −0.06 (0.04) | 0.289 | 1 | |
| Coffeescript | 0.06 (0.05) | 0.322 | 1 | |
| Javascript | 0.03 (0.03) | 0.292 | 1 | Both **IN**significant |
| Typescript | 0.15 (0.10) | − | − | |
| Ruby | −0.13 (0.05)** | <0.01 | 0.017 | |
| Php | 0.10 (0.05)* | 0.075 | 0.629 | Small Difference |
| Python | 0.08 (0.04)* | 0.075 | 0.673 | Small Difference |
| Perl | −0.12 (0.08) | 0.419 | 1 | |
| Clojure | −0.30 (0.05)*** | <0.01 | <0.01 | |
| Erlang | −0.03 (0.05) | 0.733 | 1 | |
| Haskell | −0.26 (0.06)*** | <0.01 | <0.01 | |
| Scala | −0.24 (0.05)*** | <0.01 | <0.01 | |

***$p < 0.001$, **$p < 0.01$, *$p < 0.05$

Table 6 from **CACM17** and Table 6c from **TOPLAS19**

As this is after all the additional **TOPLAS19** data gathering, cleaning, and processing, this is to a remarkable degree a confirmatory study of our **CACM19 RQ1**! In spite of that, Berger et al. have misrepresented their results as disproving our study. Interestingly, they even mis-observe values in their own table as being larger than 0.05, which affects their count of disagreements with our results[11]!

---

[11] They state in the first paragraph of **TOPLAS19**, Section 4.3: "The impact of the data cleaning, without multiple hypothesis testing, is illustrated by column (b). Grey cells indicate disagreement with the

**#miss** In spite of them showing FDR results in Table 6, as discussed above, for their conclusions they use the very conservative, and problematic Bonferoni correction[12]. They do so in a number of places, in spite of the fact that they themselves have argued against it: the **TOPLAS19** statistician author, Prof. O. Vitek, has in her prior published work consistently chosen the FDR over using Bonferroni[13]. Their own analysis includes a more appropriate p-value correction, the false discovery rate, see the FDR column in Table 6c, **TOPLAS19**. Moreover the authors themselves recognize FDR as superior in the Curry On! Talk. And yet, they defer to the Bonferroni correction, and not FDR, in their final analyses in **TOPLAS19**. They end up with only 5 significant results after the Bonferroni correction.

**#opinion #future** Bootstrap: They performed an additional *bootstrap* step, an interesting robustness analysis which considers the uncertainty in the bug-fix labeling (there are issues with labeling, which we discuss below). The results, in **TOPLAS19**, Table 6e, show that the p-value for 1 of their 5 significant coefficients from the previous step became non-significant (i.e., increased to greater than 0.05), for a final tally of 4 languages with significant associations with buggy commits, vs. our 7. While bootstrap is potentially useful vis-a-vis labeling in the presence of uncertainty, unfortunately, they used the Bonferroni correction with the bootstrap. As discussed above, that correction is too conservative, with an inappropriately deleterious effect on significant findings. We posit that much of the information in the data was lost in **TOPLAS19** after the application of Bonferroni. Additionally, for the bootstrap uncertainty they used their estimates of FP and FN in our bug-fix commit labeling, which we do not find plausible, see the discussion in bullet 7. below. We did not consider uncertainty of bug-fix labeling in our analyses. In that sense, using the bootstrap approach can be considered an improvement to our study, but only if a) hypothesis correction is done appropriately, which in **TOPLAS19** it has not, and b) if the uncertainty in bug-fix labeling is properly quantified, which as we discuss in bullet 7. below, we are not convinced Berger et al. have done.

**#opinion** Berger et al. also apply a different contrasting technique, zero-sum, than the one we used, which they claim may be more appropriate in this setting, though they give no evidence for it. In **CACM17** we have justified the use of weighted contrasts (see **CACM17**, Sect. 2.5) and provided a reference. **TOPLAS19** doesn't directly compare their contrasting technique to ours. Due to the lack of objective evidence either way, we are not swayed by their argument. At best this point is debatable, if not unnecessary.

---

conclusion of the original work. As can be seen, the p-values for C, Objective-C, JavaScript, TypeScript, PHP, and Python all fall outside of the "significant" range of values, even without the multiplicity adjustment. Thus, 6 of the original 11 claims are discarded at this stage."

[12] Most practitioners recognize that Bonferoni is too conservative and not to be used in practice, especially as it yields many false negatives and can lead to unsound statistical inference in epidemiological studies, e,g.: https://www.ncbi.nlm.nih.gov/pmc/articles/PMC1112991/, and https://academic.oup.com/beheco/article/15/6/1044/206216.

[13] E.g., https://skyline.ms/wiki/home/software/Skyline/events/2015%20US%20HUPO%20Workshop/download.view?entityId=aa2abc6d-ad47-1032-966e-da202582cf3e&name=1-Intro%20Stat%20%28Vitek%29.pdfhttps://www.ncbi.nlm.nih.gov/pmc/articles/PMC3489540/ and https://www.ncbi.nlm.nih.gov/pmc/articles/PMC3489535/

Reanalysis summary: Berger et al.'s final conclusion is virtually the same as ours: whatever positive effects were found (they found 7 significant associations before the very conservative Bonferroni correction, and 4 after; we found 9) were very small or small (we used analysis of deviance for this), and thus are possibly not real and need to be taken with caution.

**Claim 8:** Commits mis-labeled as Bug-fixing commits 36% of the time, and vice-versa 11% of the time, based on Berger et al.'s own methodology. Hence, they say our results have low power to distinguish between significant vs non-significant associations of bugs and languages.

Answer: The methodology **TOPLAS19** used for this is an alternative to the one we used.
**#opinion** Ours was not shown wrong: we reported 84% precision in both **CACM17** and **FSE14** (See section 2.4, **CACM17**). Our method is automatic, which has pluses and minuses, as discussed in our paper. Theirs is based on individual opinions of people looking at commits, so some differences in judgement can be expected. In a sense, their manual labeling overhead imposes a standard experimental power vs. cost tradeoff. Furthermore, their manual annotation protocol (viz., the instructions given to their human annotators) is not described in sufficient detail in TOPLAS19 to be repeated; it's not clear whether the human annotators were given just the commit logs? The full diffs? Access to the repos? We therefore undertook our own manual re-examination of the summary results from their manual examination, available in their artifact. When examining a subset (numbering 12)[14] of the buggy commits which they claim are false positives, considering the commit logs, the actual changes, linked issue numbers (when available) and discussions (when available) we found that 1 was a false positive, but 10 were in fact true positives, with one other one being debatable. We are therefore skeptical of their claimed 36% FP rate of our method, and look forward to the development of a clarified, repeatable protocol which could lead to more reliable results.

---

[14]1. https://github.com/GeertJohan/gorp/commit/af8337d4a1d0d35911c3cd1a8d4be58bf518fa03
2. https://github.com/sinclairzx81/typescript.api/commit/654256c47a45764f5356c71b9fd00ae8ae8a897f
3. https://github.com/mpeltonen/sbt-idea/commit/0395e37d2cde8732ca09b6f53e4313e88f3b3c22
4. https://github.com/mpeltonen/sbt-idea/commit/6a707e2887bb385a49d6ddd5d1991f95ff4675e4
5. https://github.com/MerlinDMC/gocode/commit/f36ed6ec9caf15cc7cf7fe8ec8d631ee34748d97
6. https://github.com/MerlinDMC/gocode/commit/bd72d4bc5941f538c3f789fb9a3d54a43043800b
7. https://github.com/MerlinDMC/gocode/commit/1863540842c9a5bb532d8fd75de6ede6cf5068ac
8. https://github.com/lfe/lfe/commit/e163397cdf515d653060758d8c8ae593e0ce9104
9. https://github.com/lfe/lfe/commit/14180d6f839760ce071fb49d07f21dfaaa612795
10. https://github.com/faylang/fay/commit/712bfd41d426c767a5a413ecc24911e93bcf0b54

### debatable:
11. https://github.com/Arcank/nimbus/commit/620d9580c9e401168996b17ea00e2dc77e43e245 ("fix up vertical alignment of strings---perhaps display/visualization thing, some might consider it a bug")

**Claim 9:** Finally (to end on a comic note): Dr. J. Vitek, in his many talks, publicly mocks us for including a *16-line* Perl project in a table of the "Largest Projects" in Github. Hilarity understandably ensues from the audience.

Answer:  That would indeed be a laughable error, if we had done that. Three points here: first, the table, as described in our paper, shows the *most-starred* projects  (**_not_** the *largest* ones). Second, at the time of study, the Perl file mysqltuner.pl had 784 lines of code. Finally, that particular highly-starred Perl project got filtered out of our analyzed subset, for having insufficient commit history.

**Conclusion**

We believe **TOPLAS19** confirmed most of our findings. As such it could have been a solid reproduction study of our **CACM17** paper. They proposed some improvements to our data gathering and pre-processing, proposed an alternative to our language classification (a debatable one), proposed an alternative method for labeling commits as bug-fix commits (a time consuming, and impractical for large data sets), and proposed bootstrap as an additional step for estimating significant associations between languages and bugs (interesting future work, but hampered by the overly conservative Bonferroni correction used).

Unfortunately the authors decided to position their written presentation, and (even more so) subsequent oral presentations in scholarly settings, as antagonistic to, and disproving of our work to such an extent that their and our results are not precisely and correctly represented.

Nevertheless, we sincerely believe that replications such as Berger et al. are highly valuable, even indispensable component of empirical studies in complex settings such as the social, distributed development of software, and look forward to further discussions with a more collegial spirit of rigorous scholarship.